\input harvmac
\noblackbox
\Title{}{Atmospheric Neutrino Constraints on Lorentz Violation}
 

\centerline{Sheldon Lee Glashow \  {\tt slg@bu.edu}}\bigskip
\centerline{Physics Department, Boston University}
\centerline{590 Commonwealth Avenue}\centerline{ Boston, MA 02215}
\vskip .4in

\noindent 
Sensitive tests of Lorentz invariance can emerge from the study of neutrino
oscillations, particularly for atmospheric neutrinos where the effect is
conveniently near-maximal and has been observed over a wide range of energies.
We assume these oscillations to be described in terms of two neutrinos with
different masses and (possibly) different maximal attainable velocities
(MAVs). It suffices to examine limiting cases in which neutrino velocity
eigenstates coincide with either their {\it mass\/} or {\it flavor\/}
eigenstates.  We display the modified $\nu_\mu$-$\nu_\tau$ transition
probability for each case. Data on atmospheric neutrino oscillations at the
highest observed energies and pathlengths can yield constraints on neutrino
MAV differences ({\it i.e.,} tests of special relativity) more restrictive
than any that have been obtained to date on analogous Lorentz-violating
parameters in other sectors of particle physics.

\Date{07/2004}

It seems  unlikely that Lorentz violation {\it per se,} 
rather than neutrino mass, can explain  observed
 neutrino oscillations. In particular,
atmospheric neutrinos such as are observed  at SuperKamiokande
and MACRO,  as well as K2K data, appear to be
 well described by  nearly maximal conventional ({\it i.e.,} mass-associated)
 two-flavor $\nu_\mu-\nu_\tau$ oscillations.~\ref\rex{{\it E.g.,} 
The SuperKamiokande
Collaboration, {\tt hep-ex/0404034}\semi
A. Giacomelli, A. Margiotta, {\tt hep-ex/0406037}\semi
The K2K Collaboration, 
{\tt http://neutrino.kek.jp/news/2004.06.10/index-e.htm}.}\ 
where the transition  probability of muon neutrinos 
or antineutrinos (neglecting matter effects)
is given  by:
\eqn\eone{P\big(\nu_\mu(\overline{\nu}_\mu)\rightarrow 
\nu_\tau(\overline{\nu}_\tau)\big)
\simeq 
\sin^2{\big(\delta m^2L/4E\big)},}
with $\delta m^2\simeq 2\times 10^{-3}\;\rm eV^2$.
The 2-family approximation suffices because of  the near degeneracy of two 
neutrino masses  and the smallness  of the subdominant PMNS angle
$\theta_{13}$.

 Here we ask whether small Lorentz-violating
effects involving  atmospheric neutrinos can reveal themselves as   
departures from Eq.\eone. 
We retain the  two-family formalism, presuming 
that any  MAV difference between the neutrinos
relevant to  solar neutrino oscillations plays no
significant role for atmospheric neutrinos.
 We also assume CPT conservation.
To determine the modified neutrino transition probabilities,
we  recall  Eqs.$\; 11$ and 12 of  our 1997
paper\footnote{$^*$}{Several subsequent  studies of Lorentz 
and CPT violation in the
neutrino sector~\ref\rk{A. Kostelecy,
M. Mewes, {\tt hep-ph/0308300, 0406255}\semi
A. Kostelecky, {\tt hep-ph/0403088}.}\ 
offer relevant commentary, although they are largely directed toward 
other  issues.}~\ref\rsc{S.R. Coleman, S.L. Glashow, Phys. Lett. 
B403:229,1997.},
with  $\theta_m=\theta_{23}=\pi/4$  so as to reproduce the observations
of maximal atmospheric neutrino oscillations
at low energy. 
We obtain:
\eqn\egen{P\big(\nu_\mu(\overline{\nu}_\mu)\rightarrow 
\nu_\tau(\overline{\nu}_\tau)\big)
= 
\sin^2{2\Theta}\,\sin^2{\big(\Delta L/4E\big)},}
where
\eqna\egenb
$$\eqalignno{\Delta(E)\sin{2\Theta(E)}&=\big|\delta m^2 +
2ae^{i\eta}E^2\sin{2\theta_v}\big|\,,
&\egenb a\cr
\Delta(E)\cos{2\Theta(E)} &=2aE^2\cos{2\theta_v}. &\egenb b\cr}$$
The tiny positive parameter $a$ is the 
fractional difference between the maximal attainable velocities (MAVs)
of the two neutrinos;
the phase $\eta$ is unconstrained; the angle $\theta_v$ determines the
neutrino velocity eigenstates. To interpret  this result, I define
the  critical neutrino energy:
\eqn\ecrit{E_c\equiv  \sqrt{{\delta m^2\over 2a}}\,.}
For $E\ll E_c$, Lorentz violation is ineffective. 
Oscillations are maximal with their  conventional   phase
 $\delta m^2 L/4E$.    For $E> E_c$, the neutrino mass difference
becomes  ineffective. The oscillation amplitude
is no longer maximal but   approaches $\sin^2{2\theta_v}$ as its 
phase approaches  $aEL/2$.

It is sufficient to examine the precise consequences of Eq.\egen\ in
 two  extreme  limits. 
For Case A we set $\theta_v=0$, thus taking  as the neutrino states 
with definite (and different) 
MAVs the {\it flavor\/} eigenstates, $\nu_\mu$ and $\nu_\tau$. 
  The 
conventional oscillation formula Eq.\eone\ is changed as follows:
\eqn\etwo{P\big(\nu_\mu(\overline{\nu}_\mu)\rightarrow 
\nu_\tau(\overline{\nu}_\tau)\big)
=  {1\over f^2(E)}\sin^2{\big(f(E)\,\delta m^2L/4E\big)},}
where
\eqn\ethree{f(E) \equiv \sqrt{1+4a^2\,E^4/(\delta m^2)^2}
=\sqrt{1 + (E/E_c)^4}.}
At  energies above  the critical energy
$E_c$,  oscillations  rapidly
decline in amplitude while decreasing in oscillation length.  
Loosely speaking,
oscillations remain  maximal or nearly maximal at energies below $E_c$, but 
they wash out above $E_c$.

For Case B
 we set $\theta_v=\pi/4$, thus  taking  the neutrino states with definite
MAVs to be their  {\it mass\/} eigenstates.  The neutrino oscillation
probability becomes: 
\eqn\etwob{
P\big(\nu_\mu(\overline{\nu}_\mu)\rightarrow 
\nu_\tau(\overline{\nu}_\tau)\big)
= \sin^2{\big(g(E)\,\delta m^2L/4E\big)},}
where
\eqn\ethreeb{g(E)\equiv \big| 1+2ae^{i\eta} E^2/\delta m^2\big|
= \big| 1+ e^{i\eta}(E/E_c)^2\big|.}
The oscillations remain maximal at all energies,
but their   phase increases rapidly as neutrino  energies exceed
$E_c$.

 Eqs. \etwo\ and \etwob\ show that the most sensitive tests of
Lorentz invariance may be  obtained from atmospheric neutrino observations
at the highest observed energies and longest baselines,
{\it i.e.,} at $E\sim 100$~GeV and $L\sim  10^4$~km.
 We suspect that in Cases A and B
  --- and indeed,  for any intermediate situation
described by Eq.\egen --- limits 
as severe as $E_c>100$~GeV, or equivalently
 $|a|<10^{-25}$,  can be set  through  dedicated  analyses of
currently available  Super-K or MACRO data.
  These tests of special relativity
should be of wide interest because they are considerably  more restrictive 
than any that have been reported
 on  analogous 
Lorentz-violating parameters (MAV differences) 
in other sectors of particle physics.\footnote{$^\dagger$}{See, for example,  
the constraints   summarized
in  \ref\rsg{S.R. Coleman, S.L. Glashow,
 Phys.Rev.D59:116008,1999.}, and 
more recently,  in~\ref\rgm{O. Gagnon, G.D. Moore,
{\tt hep-ph/0404196}.}.}

\bigskip
\bigskip\bigskip

Communications with Professors Giorgio  Giacomelli and Ed Kearns
are gratefully acknowledged. 
This research has been supported in part by the National Science Foundation
under grant number NSF-PHY-0099529.

\listrefs
\bye